\documentclass{aa}  
\usepackage{graphicx}
\usepackage{longtable}
\usepackage{txfonts}
\begin{document}
   \title{Milliarcsecond angular resolution of
   reddened stellar sources in the vicinity of the 
Galactic Center
\thanks{Based on observations made with ESO telescopes at Paranal Observatory}
}

\subtitle{II. Additional observations}
\titlerunning{Milliarcsecond angular resolution of
   reddened stellar sources. II}

   \author{A. Richichi\inst{1}
          \and
          O. Fors\inst{2}\fnmsep\inst{3}
          \and
          E. Mason\inst{4}
          }

   \offprints{A. Richichi}

   \institute{
             European Southern Observatory,
Karl-Schwarzschild-Str. 2, 85748 Garching bei M\"unchen, Germany
             \email{arichich@eso.org}
         \and
             Departament d'Astronomia i Meteorologia, Universitat de  
Barcelona, Mart\'{\i} i Franqu\'es 1, 08028 Barcelona, Spain
         \and
             Observatori Fabra, Cam\'{\i} de l'Observatori s/n, 08035 
Barcelona, Spain
   \and
             European Southern Observatory,
Santiago, Chile
             }

%   \date{Received September 15, 1996; accepted March 16, 1997}

  \abstract
  % context heading (optional)
  % {} leave it empty if necessary  
{
We present lunar occultation (LO)
observations obtained in August 2006 with the recently
demonstrated burst mode of the ISAAC instrument at the
ESO VLT. The results presented here follow the
previously reported
observations carried out in March 2006 on a similar but
unrelated set of sources.
}
  % aims heading (mandatory)
{
Interstellar extinction in the inner regions of the
galactic bulge amounts to tens of magnitudes at visual
wavelengths. As a consequence, the majority of sources
in that area are poorly studied and large numbers of
potentially interesting sources such as late-type giants
with circumstellar shells, stellar masers, infrared stars,
remain excluded from the typical investigations
which are carried out in less problematic regions.
Also undetected are a large numbers of binaries.
By observing LO events in this region, we gain the means
to investigate at least a selected number of sources
with an unprecedented combination of sensitivity and
angular resolution.
}
  % methods heading (mandatory)
{
The LO technique permits to achieve milliarcsecond resolution
with a sensitivity of K$\approx$12\,mag
at a very large telescope. We have used the opportunity of
a favorable passage of the Moon over a crowded region in the
general direction of the Galactic Center
to observe 78 LO events of heavily reddened stellar sources.
}
  % results heading (mandatory)
{We have detected six new binary and one triple star, with typical
projected separation of $\approx 0\farcs01$. We have also detected
the compact circumstellar emission around one maser and one
central star of a planetary nebula. Additionally we have measured
the diameter and/or circumstellar shell of two carbon stars
and other IR sources.
}
  % conclusions heading (optional), leave it empty if necessary 
{We have used the upper limits on the size of about 60 unresolved
or marginally resolved sources to establish the performance of the
method. In agreement with our previous result, we conclude that
lunar occultations in fast read-out mode on a detector subwindow
at an 8\,m-class telescope can achieve an angular resolution close
to $0\farcs001$ with a sensitivity K$\approx 12$\,mag.
}

   \maketitle
%
%________________________________________________________________

\section{Introduction}
This paper follows closely the scientific rationale,
method, observations and results
already presented 
in Richichi et al. (\cite{PaperI}, Paper~I hereafter),
where we introduced the technique of lunar occultations (LO) at the
ESO Very Large Telescope using a fast readout (burst mode) of the
ISAAC instrument. For our application, 
the burst mode allows to sample subarrays of 32 or 64 squared pixels
with sampling times of 3.2 or 6.4\,ms, respectively.
We demonstrated that in this way it is possible to record LO data
with the highest quality ever achieved 
in terms of signal-to-noise ratio (SNR),
thanks also to the strong reduction of scintillation made possible
by the 8.2\,m mirror in the near-IR. In Paper~I
we characterized in detail 
the performance, which approached $\approx 1$
milliarcsecond (mas) in angular resolution, K$\approx 12$\,mag in
sensitivity, and $\Delta$K$\approx 8$\,mag in dynamic range on scales
smaller than the Airy disk of the telescope.
Notwithstanding the obvious limitations of LO with respect to the
choice of the sources and the repeatability, this combined performance
is superior to that of any other technique presently available
for high angular resolution in the near-IR.

In Paper~I we took advantage of a recently developed
data pipeline which employs
an automated generation of masks and light curve extraction  and
generates  the first-guess parameters for the fitting
(AWLORP, Fors et al. \cite{fors08}). This pipeline allows us to
perform a quick preliminary inspection of large volume of LO data.
Subsequently,
a more detailed interactive analysis is performed on a selected
number of sources using both
model-dependent and model-independent procedures
(ALOR and CAL respectively,
Richichi et al. \cite{richichi96}
and
Richichi \cite{CAL}).

In the present paper we report on LO observations
carried out in the night of August 5, 2006,
during which 78 events were observed. 
We descrive the observations and the list of sources
in Sect.~\ref{data}, without 
recalling the details of the method and of the
data analysis which 
can be found in Paper~I and references therein.
Subsequently, 
we discuss the results in Sect.~\ref{results}.
These include
several new detections of binary and triple stars,
as well as angular diameters and circumstellar shells of
diverse objects such as a maser, two carbon stars, and the
central star of a planetary nebula.

%__________________________________________________________________

\section{Observations}\label{data}
\begin{figure*}
\includegraphics[width=18cm]{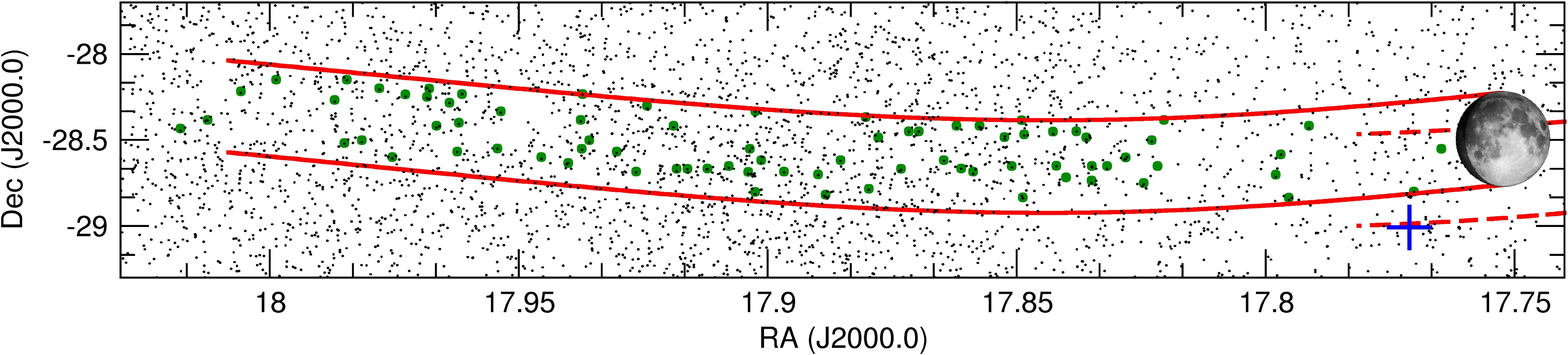}
\caption {The region close to the sources occulted by the
Moon in the night of August 5, 2006, as seen from Paranal. 
The small dots represent
sources in the 2MASS catalogue with magnitude K$\le 7.5$. The heavier
dots are the sources for which we could record an occultation.
Also shown by the solid lines is
the apparent path of the Moon, moving from West to East.
Part of the path of March 21, 2006 (Paper~I) is shown by the
dashed lines.
The cross indicates the Galactic Center.
}
\label{fig_location}
\end{figure*}

\begin{figure}%f1
\includegraphics[width=8.8cm]{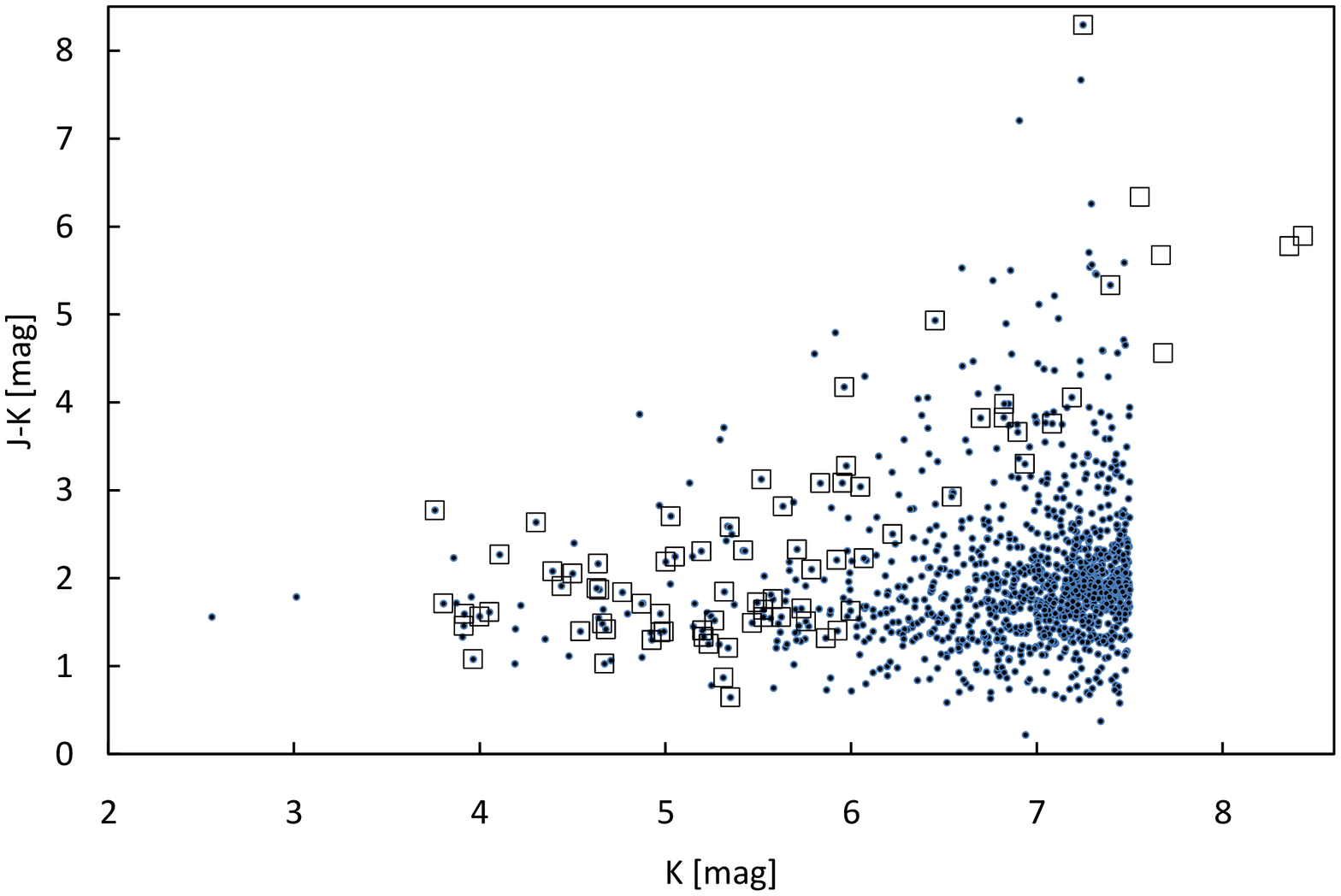}
\caption {Color-magnitude diagram for the sources 
of which we recorded LO events as 
listed in Table~\ref{lo_complete} (open squares).
The small dots mark all sources occulted by the Moon
during the same night. For ease of presentation we show them
only to the limit K$\le 7.5$\,mag.
}
\label{fig_km}
\end{figure}

We observed a passage of the Moon in a crowded, heavily reddened
region in the direction of the
Galactic Bulge, in the night of August 5-6, 2006. The center
of this region was located at $\approx 17^{\rm h} 53^{\rm m}$
and $-28\degr31\arcmin$. Fig.~\ref{fig_location} shows the
area, with the apparent lunar path superimposed.
Note that this region is contiguous but barely overlapping
with the area of the events
reported in Paper~I. In particular, there were no sources
observed on both dates.
Although the minimum approach of the Moon to the Galactic Center,
as seen from Paranal, was larger in August than in March 2006
(12$\arcmin$ and 12$\arcsec$ respectively), the region
is nevertheless 
heavily reddened by interstellar dust. A near-IR color-mgnitude
diagram is shown in Fig.~\ref{fig_km}, where we include only sources
with K$\le 7.5$\,mag to avoid overcrowding. 
% Figure 1 available both printed and electronically

Our predictions were based on the 2MASS Catalogue
(Cutri et al \cite{2003yCat.2246....0C}).
To the limit K$\le8.5$\,mag
a total of 6421 sources were due to be
occulted during the night. 
As was the case in Paper~I, it would have been impossible
to observe all these events, not to mention the even huger number
of those possible up to the
sensitivity limit of the technique. We then applied a prioritization
scheme which took into account
the magnitude, 
the color (redder objects having higher priority), 
and the time intervals between the events. Priorities were 
increased for sources with known cross-identifications and
previous measurements.
We carried out our observations
over about 8.5 hours,
during which a total of 78 events 
were recorded. 
We took advantage of the experience with such observations
from our previous run, 
and of the fact that the August
events were disappearances (lunar phase 84\%), making the
pointing significantly easier. 
Also the image quality was significantly improved, with frequent
checks of the active optics correction for the primary mirror.
Our main limitation
were, as before, the overheads for 
telescope pointing and data storing. At best, we could
record an occultation every three minutes.
Given the more favorable circumstances in this second
run, we adopted a subwindow
size of 32x32 pixels for most of the events, and detector
integration time (DIT) of 3.2\,ms. This is also the
sampling time. In a few cases, we adopted
the configuration 64x64 pixels and DIT=6.4\,ms. 
The field of view in the two cases was
about $4\farcs7$ and $9\farcs5$ squared.
Data acquisition was triggered automatically according to the
time of the event, and the typical 
length of the recorded data was between 10 and 20
seconds, in order to account for some uncertainty in the predictions.

%______________________________________________________________
% In the text, at the place where the large table should appear
% add the command:
% counter for longtable at the end of the document: List of observed LO events
\addtocounter{table}{1}
% Tables counters will be well numbered.
%------------------------------------------------------------------------------
\section{Results}\label{results}
A list of the sources observed, including details of observation
and comments, is provided in Table~\ref{lo_complete}.
We list for completeness  the total of
78 recorded events, but in 6 cases no LO was detected.
Most of the sources have no entries in the Simbad database,
with only 10 of the sources in
Table~\ref{lo_complete} having a cross-identification, and only
4 having a spectral determination.
The K magnitudes range from 3.8 to 8.4\,mag, a wider range
than in Paper~I, and the J-K colors from 0.6 to 8.3.
The complete set of light curves and best fits
is available online.
In the following
we will discuss individually the sources which were found to
be resolved or binaries, 
and separately we will use the unresolved sources
to characterize the performance of the method.

Table~\ref{tab:results}
lists sources which were found to be
resolved in our observations, either with an angular diameter
or with a compact circumstellar shell, or as binary stars. 
The table
uses the same format already adopted in Paper~I and in previous
papers referenced therein.
In summary, the columns list the value of the fitted linear rate of
the event V, its deviation from the predicted rate V$_{\rm{t}}$,
the local lunar limb slope $\psi$, the position and
contact angles, and the SNR. For binary detections, the
projected separation and the brightness ratio are given, while for resolved
stars
the angular diameter $\phi_{\rm UD}$ is reported, under the assumption of
a uniform stellar disc, or the characteristic size of the shell.
All angular quantities are computed from the fitted
rate of the event.
\begin{table*}
\caption{Summary of results.}
\label{tab:results}
\centering          
%\scriptsize
\begin{tabular}{lcrrrrrrrr}
\hline 
\hline 
\multicolumn{1}{c}{(1)}&
\multicolumn{1}{c}{(2)}&
\multicolumn{1}{c}{(3)}&
\multicolumn{1}{c}{(4)}&
\multicolumn{1}{c}{(5)}&
\multicolumn{1}{c}{(6)}&
\multicolumn{1}{c}{(7)}&
\multicolumn{1}{c}{(8)}&
\multicolumn{1}{c}{(9)}&
\multicolumn{1}{c}{(10)}\\
\multicolumn{1}{c}{Source}&
%\multicolumn{1}{c}{$|$V$|$ (m/ms)}&
\multicolumn{1}{c}{V (m/ms)}&
\multicolumn{1}{c}{V/V$_{\rm{t}}$--1}&
\multicolumn{1}{c}{$\psi $($\degr$)}&
\multicolumn{1}{c}{PA($\degr$)}&
\multicolumn{1}{c}{CA($\degr$)}&
\multicolumn{1}{c}{SNR}&
\multicolumn{1}{c}{Sep. (mas)}&
\multicolumn{1}{c}{Br. Ratio}&
\multicolumn{1}{c}{$\phi_{\rm UD}$ (mas)}\\
\hline 
17512677-2825371 & 0.2787 & $-$24.9\% & $-$10 & 25 & $-$64 & 66.5 &    &    & \multicolumn{1}{c}{13.2 (shell)}   \\
17514339-2825469 & 0.3825 & $-$2.9\% & $-$1 & 35 & $-$52 & 104.6 & 7.4 $\pm$ 1.5 & 6.1 $\pm$ 0.3 &    \\
17524687-2847207A-B & 0.5291 & 5.0\% & 5 & 130 & 45 & 99.6 & 8.7 $\pm$ 1.1 & 11.8 $\pm$ 1.2 &    \\
17524687-2847207A-C & & & & & & 93.1 & 6.5 $\pm$ 0.2 & 73 $\pm$ 29 &    \\
17524903-2822586 & 0.3097 & $-$10.7\% & $-$4 & 22 & $-$61 & 186.7 & 13.1 $\pm$ 2.4 & 31.2 $\pm$ 2.5 &    \\
17531817-2849492 & 0.3932 & 14.0\% & 5 & 147 & 64 & 127.1 &    &    & 5.83 $\pm$ 0.10 \\
17534818-2841185 & 0.6422 & 4.3\% & 7 & 113 & 29 & 213.1 & 112 $\pm$ 10 & 121.4 $\pm$ 19.8 &    \\
17540891-2820125 & 0.3044 & $-$12.0\% & $-$4 & 19 & $-$62 & 160.0 & 7.56 $\pm$ 0.16 & 7.38 $\pm$ 0.15 & 5.30 $\pm$ 0.06 \\
17545806-2840144 & 0.6266 & 0.9\% & 1 & 111 & 30 & 339.6 &    &    & 2.73 $\pm$ 0.13 \\
17553507-2841150 & 0.6060 & 7.2\% & 5 & 125 & 45 & 149.3 & 7.36 $\pm$ 0.34 & 6.05 $\pm$ 0.35 & \multicolumn{1}{c}{shell?}   \\
17582187-2814522 & 0.7391 & $-$2.8\% & $-$3 & 46 & $-$32 & 91.3 &    &    & 4.72 $\pm$ 0.49 \\
18004499-2823118 & 0.8590 & $-$1.9\% & $-$2 & 101 & 23 & 267.2 & 18.6 $\pm$ 1.2 & 77.1 $\pm$ 9.2 &    \\
\hline 
\hline 
\end{tabular}
\end{table*}

\subsection{Resolved sources}\label{resolved}
{\object 2MASS 17512677-2825371}
 is coincident with the radio-luminous maser {\object OH 1.09-0.83},
 which has no optical counterpart. 
 Jones et al. (\cite{jones88})
 used near-IR spectra to classify this star as a true core-burning
 supergiant. Our data are not easy to fit over an extended time sequence,
 due to flux oscillations which  could be due to 
 telescope jitter coupled to the small field of view
 over the extended emission around this source.
Also the fitted rate of the event is significantly different from
the predicted one (see 
Table~\ref{tab:results}), which is not out of the ordinary
however given the rather high value of the CA.
 In any case, the data around the central time of occultation unambiguously
 indicate a very resolved source.
 A CAL analysis shows a central peak with about
 11\,mas FWHM, surrounded by rather symmetrical wings extending
 to about $\pm0\farcs1$. This is in good agreement with the
 conclusions of 
 Cobb \&  Fix (\cite{cobb87}), who 
 obtained 1-D speckle observations
 of this source at 10 and 5\,$\mu$m, finding it to be generally
 symmetric with angular sizes of $\approx0\farcs3$ and 
 $<0\farcs1$, respectively.

{\object 2MASS 17531817-2849492} is identified with a carbon star
well detected both in the visual
({\object CGCS 3889}, V=11.6\,mag) and in the mid-IR
({\object IRAS 17501-2849}, 6\,Jy at 12\,$\mu$m). Our data are
best fitted by a resolved central star 
of 5.8\,mas
(see Table~\ref{tab:results}) and a circumstellar component.
The CAL analysis reveals that
this latter contributes 9.2\% of the total K-band flux,
and extends over about $0\farcs25$ with an asymmetric profile.
Two narrow maxima of emission from the shell can be detected
at $\approx \pm$10\,mas from the position of the central star, 
or about 2~R$_{*}$, and
are interpreted as the inner rim. Unfortunately, no estimates
of the distance to {\object CGCS 3889} are available.  Based
on its brightness and the relatively large angular sizes of the
star and the shell, we speculate that this source is well in the
foreground with respect to the galactic bulge and that its
J-K color of 2.6\,mag is of local origin.

{\object 2MASS 17540891-2820125} is a very reddened source,
with K=3.8\,mag from the 2MASS catalogue.  
Other designations include 
{\object IRC-30330} and {\object AFGL 5151S}.
Hansen \& Blanco (\cite{hansen73}) associated it with
an optical counterpart having V$\approx 15$\,mag
and spectral type 
M8 or cooler. Variability is of course prominent, and even in
K-band the brightness appears to have decreased by about 1.5\,mag from
the measurement in the above paper to the 2MASS value.
Our data show a significantly resolved star, and in fact our
data-independent CAL analysis has revealed a close companion
which has subsequently been confirmed in an improved
model-dependent binary fit.
A near-IR LO of this source 
was already recorded in 1988 (i.e. one Saros cycle earlier)
and discussed by Richichi et al. (\cite{richichi92}, RLD
hereafter).
Also at that time the angular diameter was resolved, albeit
with a slightly smaller diameter than our present measurement.
We have re-analyzed the older data, and looked for the presence
of a faint companion. Luckily, the data quality was sufficiently high
also in the previous observation (SNR=127),
and indeed we now confirm 
that the 1988 data are indeed better fitted with a resolved star
and a companion than by the RLD single component fit.
The combination of the position angles and projected
separations of the 1988 and 2006 data would yield the true parameters
PA=$56\fdg9\pm1\fdg7$ and $\rho = (9.6\pm0.2)$\,mas. This however is
subject to the condition that the orbital motion has been negiglible in the
intervening 18 years, which is difficult to verify in the absence of a
reliable distance to the star. We note in conclusion that, albeit the
smaller angular diameter measured in 1988 is subject to more uncertainty
than the present one due to a number of artifacts described in detail
in RLD, this is consistent at least in principle with the brightness
decrease observed since.

{\object 2MASS 17582187-2814522}  exhibits significant
reddening (J-K=3.8\,mag) and is associated with the
planetary nebula 
{\object ESO 456-27}. Cross-identifications with a
wide range of nomenclature exist, including 
{\object IRAS 17552-2814} which reports a flux of
8.0\,Jy at 12\,$\mu$m and
24,7\,Jy at 100\,$\mu$m, but only upper limits at other
wavelengths. Our data are of course sensitive only to the
inner regions on the sub-arcsecond scale, and they reveal
a compact, clearly resolved source. When fitted with a 
uniform-disk model, the angular diameter is 4.7\,mas
(see Table~\ref{tab:results}). 
Tajitsu \& Tamura (\cite{tajitsu}) have computed distances for
a long list of planetary nebulae, but they have
excluded
{\object ESO 456-27} because of its IR flux being
not consistent with a black body.
Zhang (\cite{zhang}) used different statistical relations and
estimated a distance of 8.2\,kpc, while 
Cahn et al. (\cite{cahn92}) had derived the value
5.35\,kpc.
In spite of the uncertainties, it is clear that our result
corresponds to a linear extent $\la 25$\,AU and is thus unrelated
to the size of the central star. The most likely interpretation
is that we are detecting the emission from the
dust around the central star. Unfortunately the lack of
extensive infrared fluxes prevents detailed estimates of
the spectral energy distribution and further conclusions.
To be consistent with the arguments given later 
in Sect.~\ref{sect_unres}, we note that the SNR and diameter
accuracy for this source place it at the margin of what can
be considered reliably resolved in Fig.~\ref{fig_diam}.
\begin{figure}
\includegraphics[width=8.8cm]{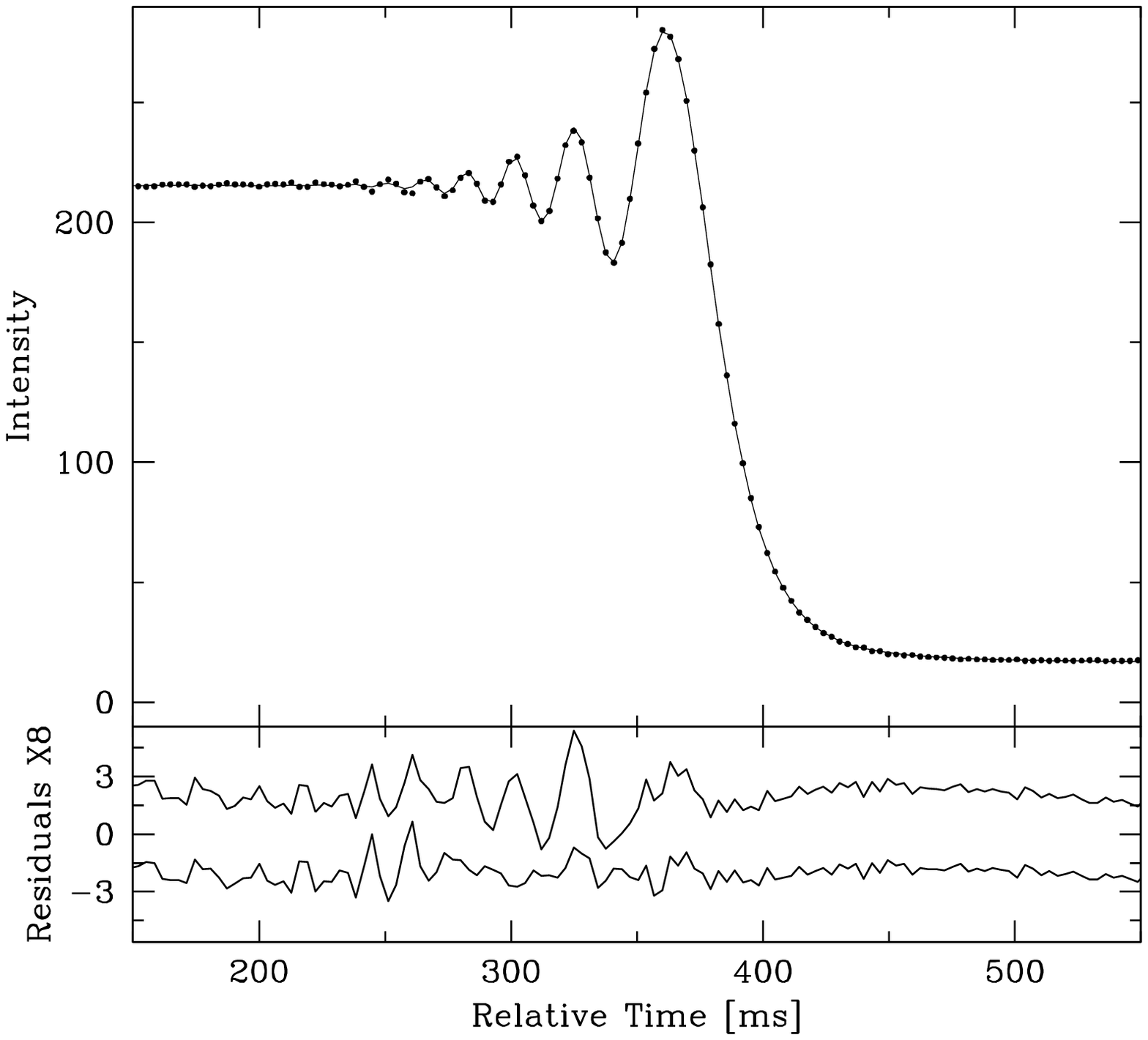}
\caption {
Top panel:
data (dots) and best fit (solid line)
for
{\object 2MASS 17545806-2840144}.  The lower panel shows, on an
enlarged scale and
displaced by arbitrary offsets
for clarity, the residuals of the fits by a point-like 
(above, reduced $\chi^2=2.9$) and 
a resolved uniform-disk  model
(below, reduced $\chi^2=1.2$) as listed in 
Table~\ref{tab:results}.
}
\label{fig_gc326}
\end{figure}

Using the same argument but with opposite conclusions, 
we note that the source
\object{2MASS 17545806-2840144} has a relatively small
diameter value but appears to be convincingly resolved
using the guidelines of Fig.~\ref{fig_diam}.
Thus we include this source in
Table~\ref{tab:results}, although the expected diameter
for a late-type star of the same brightness (K=4.4\,mag) in the
absence of circumstellar emission would be significantly
smaller than 1\,mas. In the absence of cross-identifications
and distance estimates, it is difficult to argue if
the color J-K=2.1\,mag could be due
to local or to interstellar extinction.

\subsection{Binaries}\label{sect_bin}
In addition to 
{\object IRC-30330} mentioned above, other 6
sources were found to be binary or triple,
with relatively small projected separations and magnitude
differences between the components up to
$\Delta{\rm K}\approx5$\,mag.
They are
all new detections, and for three of them no optical
or infrared counterparts are known and no literature
references exist. These include 
{\object 2MASS 17514339-2825469},
{\object 2MASS 17524687-2847207} and
{\object 2MASS 17524903-2822586}.
Their J-K colors, while being quite red, are not especially
extreme compared with the rest of the sample and are
probably indicative of a foreground location with respect to 
the bulge. Under this assumption, their projected
separations would correspond to about 10-50\,AU.

\begin{figure}
\includegraphics[width=8.8cm]{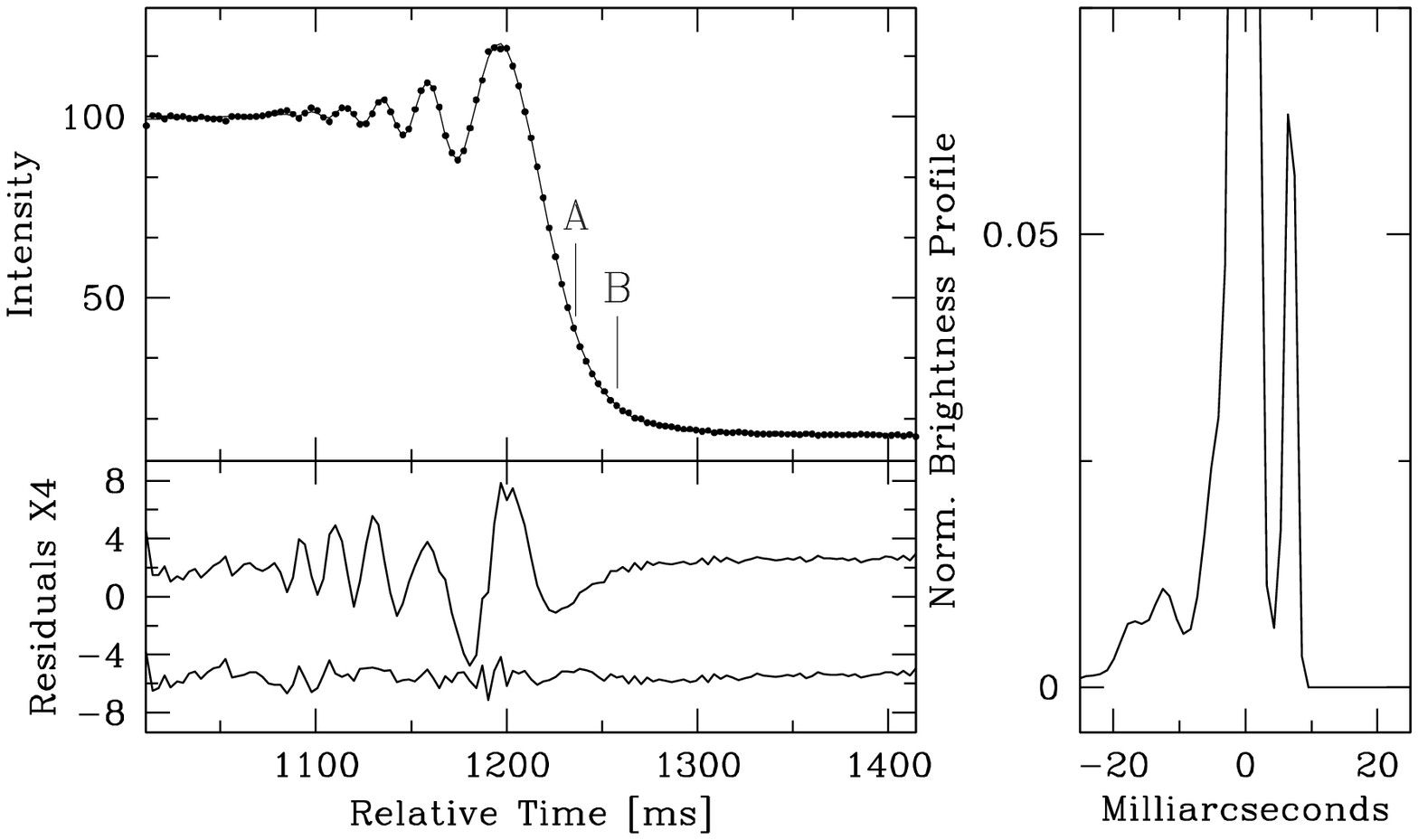}
\caption {{\it Left:} upper panel, data (dots) and best fit (solid line)
for
{\object 2MASS 17553507-2841150}.  The lower panel shows, on an
enlarged scale and
displaced by arbitrary offsets
for clarity, the residuals of the fits by a point-like 
(above)
and  by a binary star model
(below)
as listed in Table~\ref{tab:results}.
The times of the geometrical occultation
of two stars are also marked, with their difference
corresponding to a separation of 7\,mas.
{\it Right:} brightness profile reconstructed by the
model-independent CAL method.
}
\label{fig_gc333}
\end{figure}

The scenario is similar for the remaining three new binaries,
which do have known cross-identifications.
{\object 2MASS 17553507-2841150}  has no optical counterpart,
but is well detected in the mid-IR as
{\object IRAS 17524-2840}, with fluxes of 
about 23, 10 and 3 Jy at 12, 25 and 60\,$\mu$m.
Volk et al (\cite{volk91}) found the IRAS low resolution spectrum
of this source to be unusual and not easily classifiable.
On the basis of both IRAS and ground-based IR photometry
Gugliemo et al (\cite{guglielmo93}) classified it as 
a carbon star. 
Fig.~\ref{fig_gc333} shows that our data are best fitted by a
binary model.
We note that the diameter of the primary is found to be
$3.28\pm0.32$\,mas, but we consider
this as marginally resolved as discussed in Sect.~\ref{sect_unres}.
The CAL analysis shown in the right panel of
Fig.~\ref{fig_gc333} also seems to indicate some 
extended emission on the scale of a few stellar radii, but since
in this case we used a low-order Legendre polynomial to fit
some scintillation (Richichi et al \cite{richichi7s}), we are not able
to further confirm it.
We note that the brightness of the object seems to have decreased
from K=3.17\,mag reported by
Gugliemo et al (\cite{guglielmo93}) to K=4.11\,mag measured
by 2MASS. The J-K colors have changed more markedly.

{\object 2MASS 17534818-2841185} has a bright optical counterpart
in {\object HD 162761}. The magnitudes are V=7.9, K=5.3 and the
spectral type K0III. No references are found in the literature
for this star. 
{\object 2MASS 18004499-2823118}  is the infrared source
{\object IRAS 17575-2823}, with 3.1\,Jy at 12\,$\mu$m.
It is undetected in the other IRAS bands, and has no optical counterpart.
Follow up studies of all these sources
are desirable,  but they appear challenging with
any other technique other than LO.
We also note that some other stars in our sample have companions on
much larger (arcsecond) scales.  These are marked in 
Table~\ref{lo_complete}, but we do not provide details as they
can be easily detected with standard imaging.

\subsection{Unresolved sources and performance}\label{sect_unres}
\begin{figure}
\includegraphics[width=8.8cm]{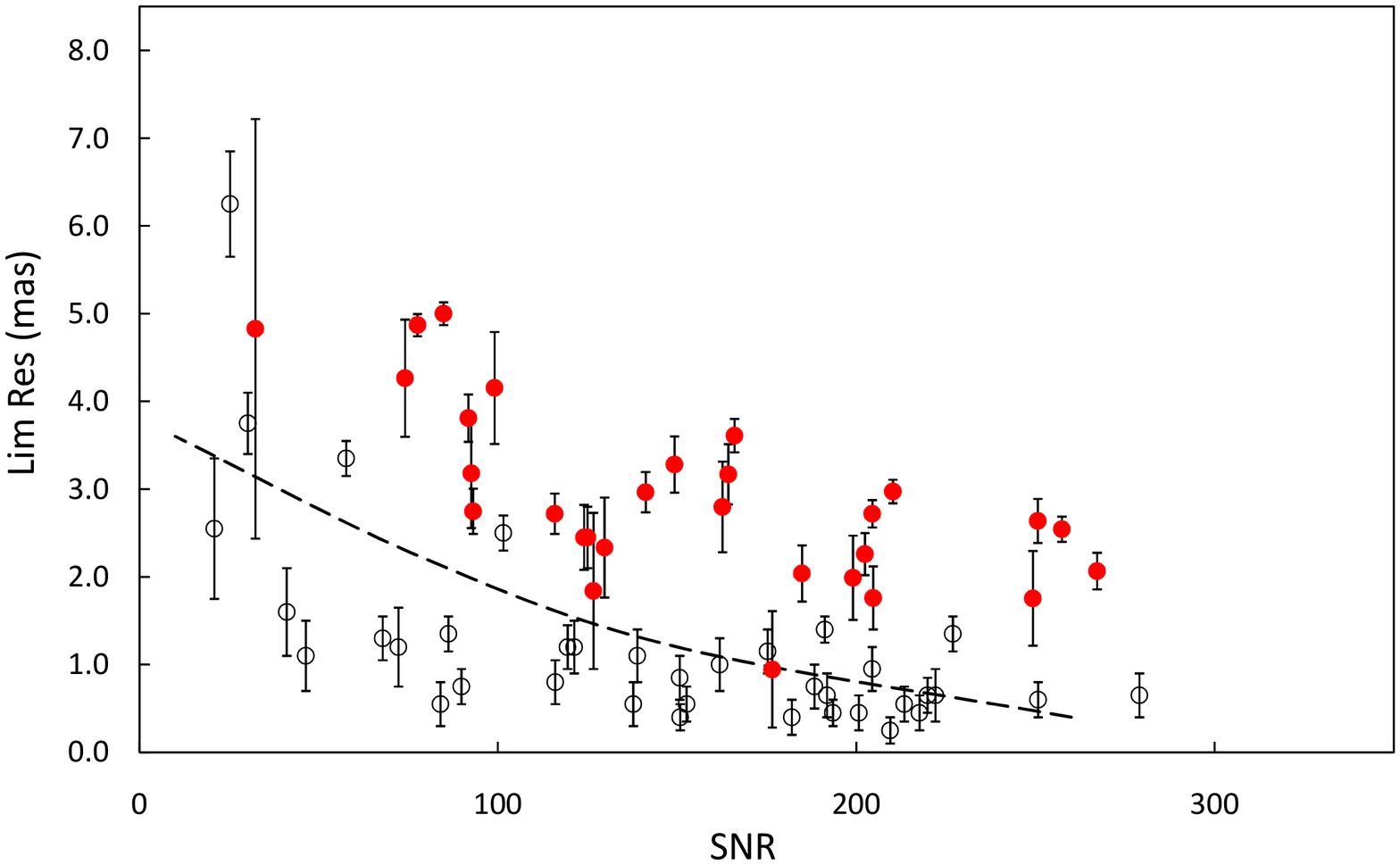}
\caption {Sources for which an upper limit on the angular diameter
could be established using the method described in the text are
marked with outlined circles. The filled circles are sources for
which an angular diameter could be formally established, but for
which the associated errorbar is such that we do not consider
them reliably resolved.
The dashed line is the arbitrary representation
of a tentative SNR-limiting angular resolution relationship
drawn in Paper~I.
}
\label{fig_diam}
\end{figure}

We have tested
the resolved/unresolved character of all the sources in our
sample using a criterium of variations in the $\chi^2$ of the fit,
as described in Paper~I and references therein.
Excluding the sources just discussed in Sect.~\ref{resolved},
the remaining 61 stars were found to be unresolved (i.e. with
an upper limit on the angular diameter), or  only
marginally resolved (i.e. with an angular diameter which is
not distinguishable from unresolved when the error bars
are taken into account).
We plot these values as a function
of SNR in Fig.~\ref{fig_diam}, where it can be seen that
the empirical relationship for the limiting angular
resolution tentatively established in Paper~I
still holds satisfactorily. 

\begin{figure}
\includegraphics[width=8.8cm]{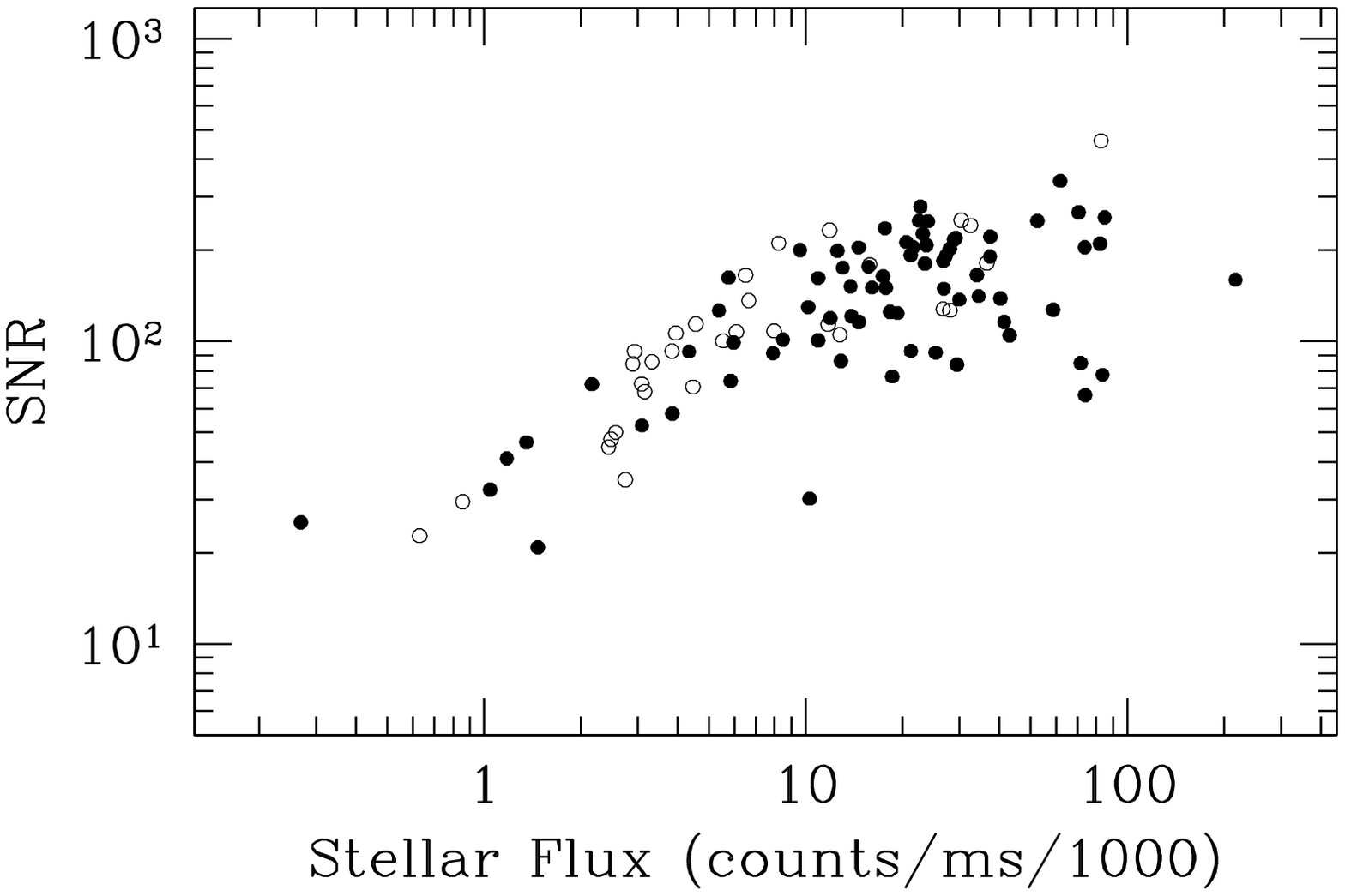}
\caption {
Plot of the SNR achieved in our fits of both resolved and
unresolved source, as a function of the measured flux
(solid points). The outlined points are the same from
Paper~I.
}
\label{fig_k0snr}
\end{figure}

Also the relationship between stellar flux and SNR is in
good
agreement with the one previously reported in Paper~I,
as shown in Fig.~\ref{fig_k0snr}. The longer DIT employed
in the observations of Paper~I does not seem to have
clear-cut consequences: on one side more photons are collected
for the same magnitude, but on the other side the
time sampling of the light curve is poorer.
As before, we estimate the limiting magnitude of the method
at K$\approx 12$\,mag for SNR=1.
However,
it can be noticed that the points relative to the August 2006
observations tend to have more scatter, especially at the
fainter magnitudes.

\begin{figure}
\includegraphics[width=8.8cm]{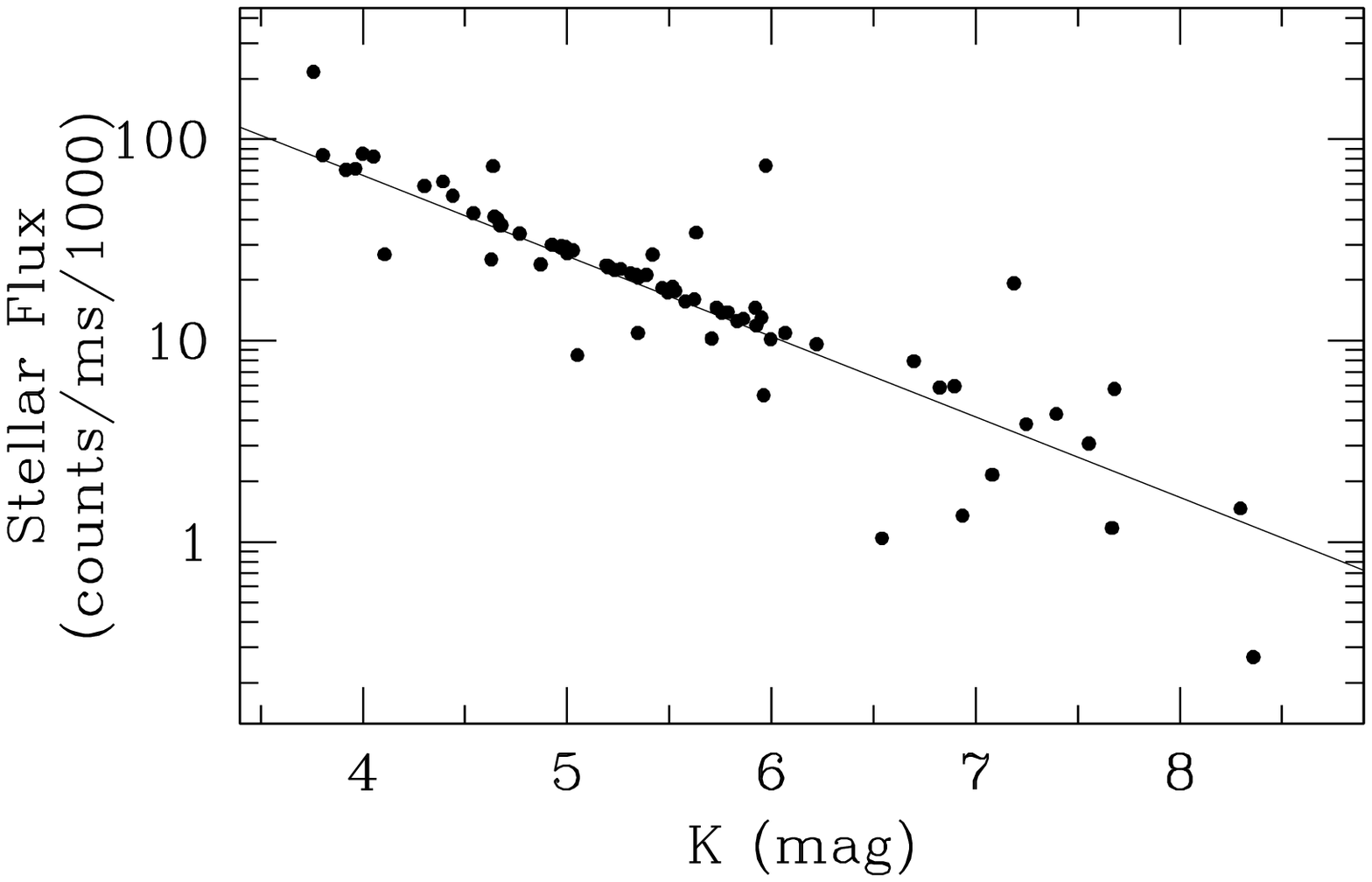}
\caption {
Plot of the measured flux against the 2MASS K magnitude of the
sources. The line shows the expected counts, based on
the ISAAC Exposure Time Calculator. 
}
\label{fig_k0f0}
\end{figure}

We believe that this scatter is due to the fact that the
August run included more faint sources than the March one,
and also to intrinsic variability. We show 
in Fig.~\ref{fig_k0f0} a plot of the measured flux against
the 2MASS magnitude.
In the absence of the image clipping and image quality
problems encountered in Paper~I, the distribution of the
points in Fig.~\ref{fig_k0f0} follows very closely the
expected ISAAC performance. 

It can now be clearly seen that several sources had a flux,
at the time of our measurement, significantly different from
the 2MASS value. This is indicative of variability, both
as brightening and as dimming, of up to two magnitudes in the
K band.
Not surprisingly, some of our well resolved sources with
a complex structure are also those which exhibit the
largest difference between our measured flux 
and the 2MASS magnitude. These include
the maser
{\object 2MASS 17512677-2825371}, 
almost six times brighter in August 2006 than at the time
of the 2MASS measurement, and {\object IRC-30330},
1.6 times brighter.
We also mention that 
{\object 2MASS 17545806-2840144} appeared significantly
brighter too, which could help in understanding why the
source appears larger in size than expected.
The putative carbon star
{\object 2MASS 17553507-2841150}, for which we detected
a new companion, was almost two times fainter 
at the time of our measurement
than expected.

%------------------------------------------------------------------------------
% Figure electronically only
\onlfig{8}{
\begin{figure*}%f4
\resizebox{\hsize}{!}{\includegraphics{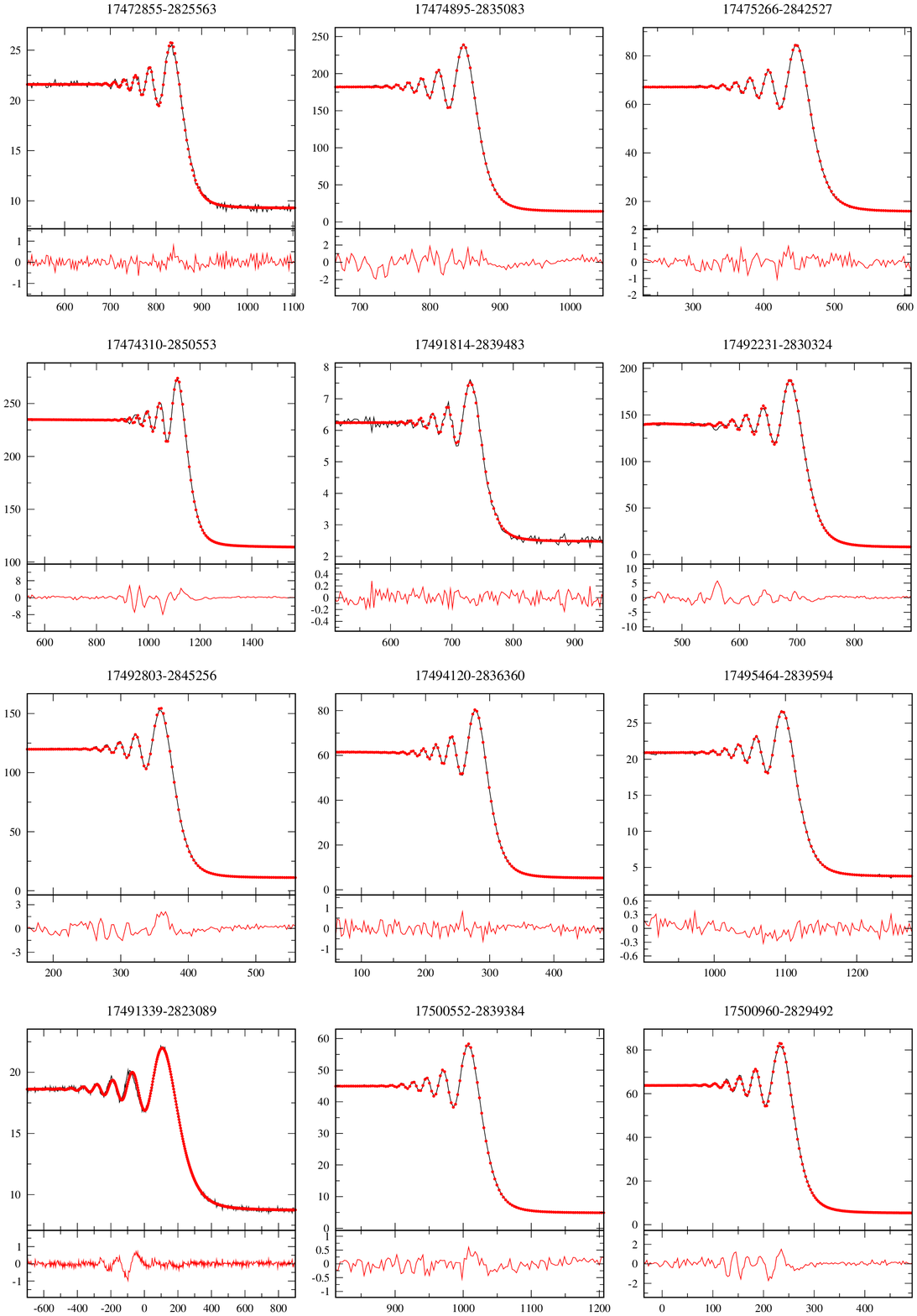}}
\caption {Top panels: LO data and
best fit. Bottom panels: fit residuals on an enlarged scale.}
\label{fig:LO}
\end{figure*}
}

%------------------------------------------------------------------------------
\onlfig{8}{
\begin{figure*}%f4
\resizebox{\hsize}{!}{\includegraphics{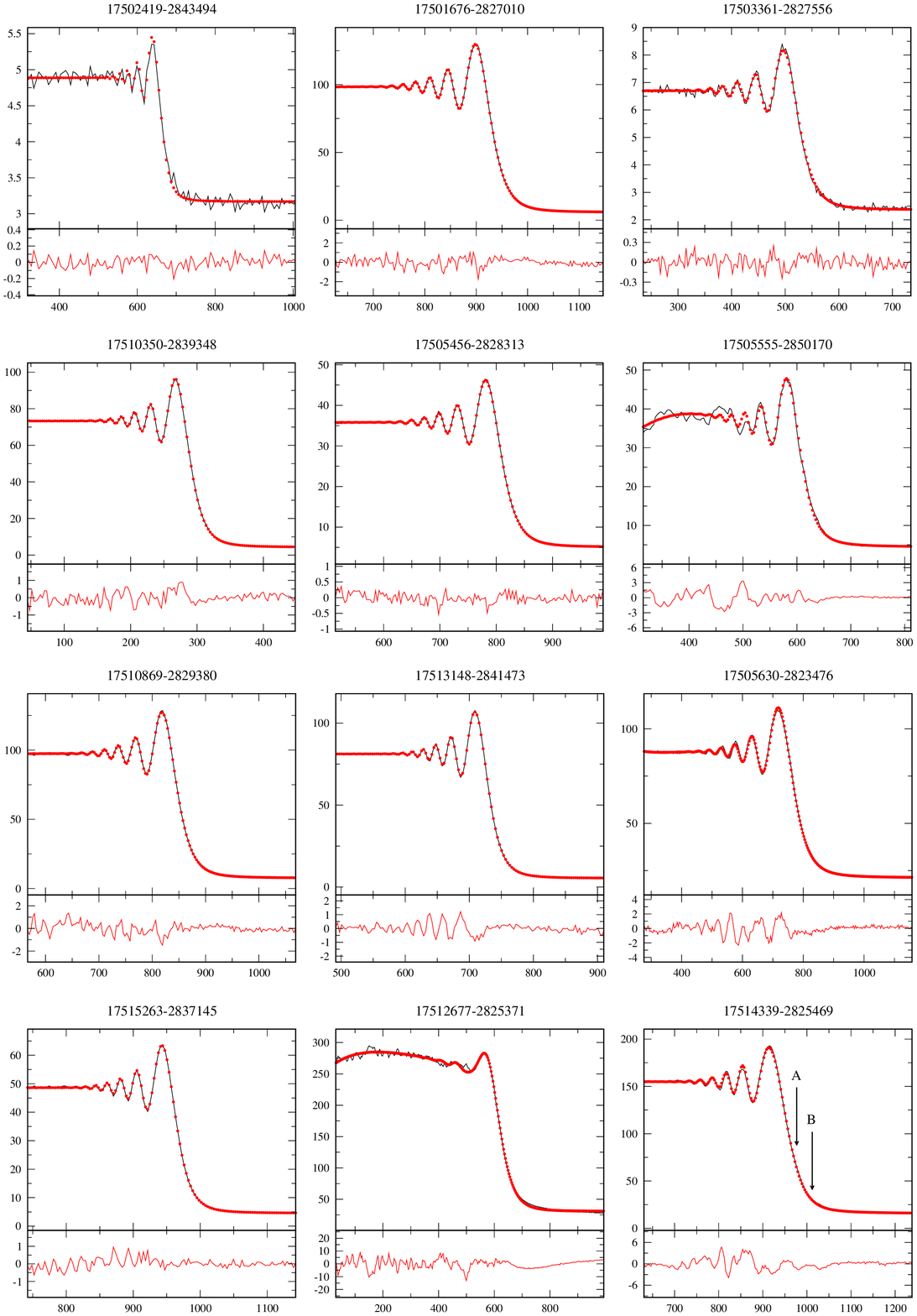}}
\caption {Top panels: LO data and
best fit. Bottom panels: fit residuals on an enlarged scale
(continued).}
\end{figure*}
}

%------------------------------------------------------------------------------
\onlfig{8}{
\begin{figure*}%f4
\resizebox{\hsize}{!}{\includegraphics{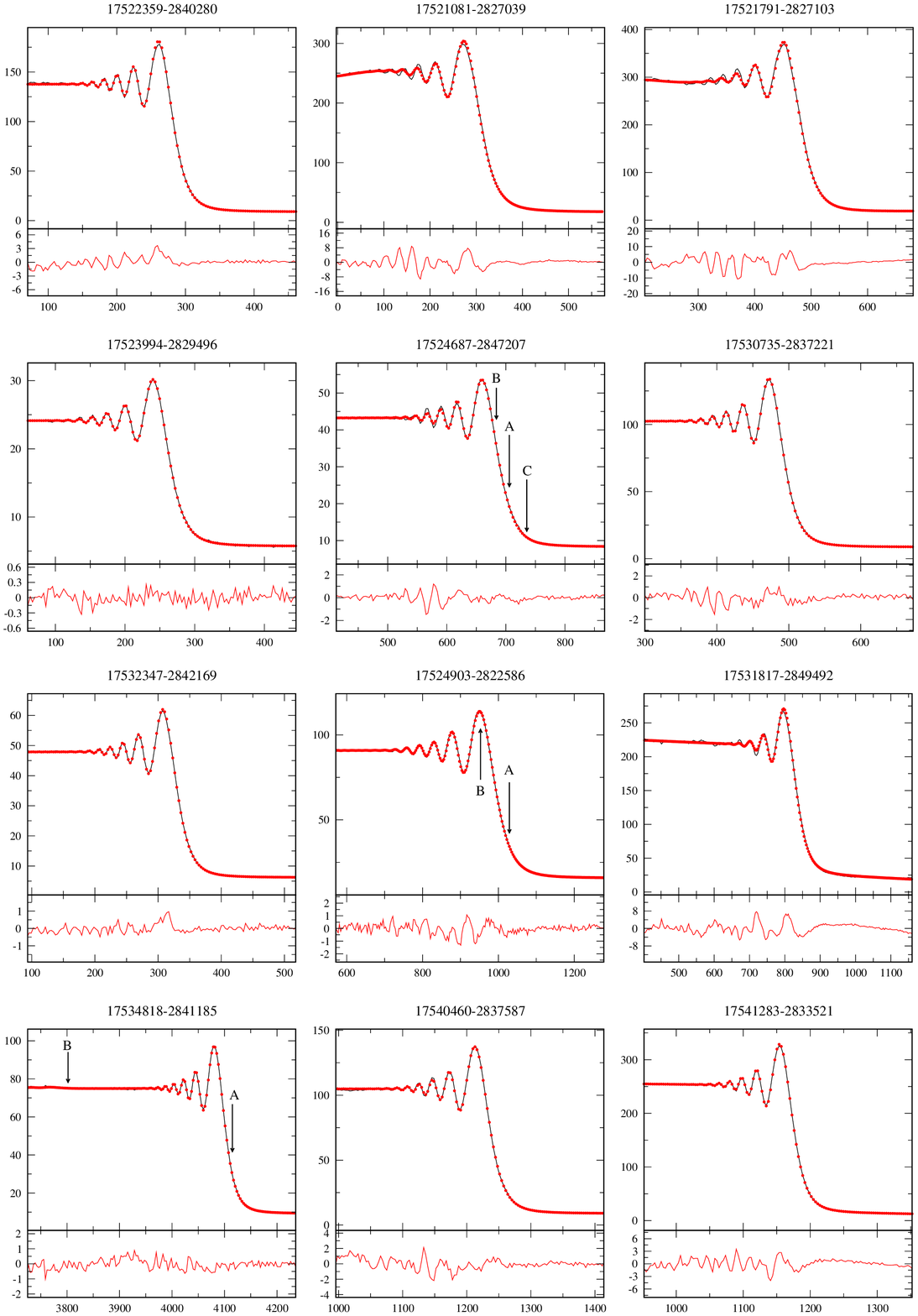}}
\caption {Top panels: LO data and
best fit. Bottom panels: fit residuals on an enlarged scale
(continued).}
\end{figure*}
}

%------------------------------------------------------------------------------
\onlfig{8}{
\begin{figure*}%f4
\resizebox{\hsize}{!}{\includegraphics{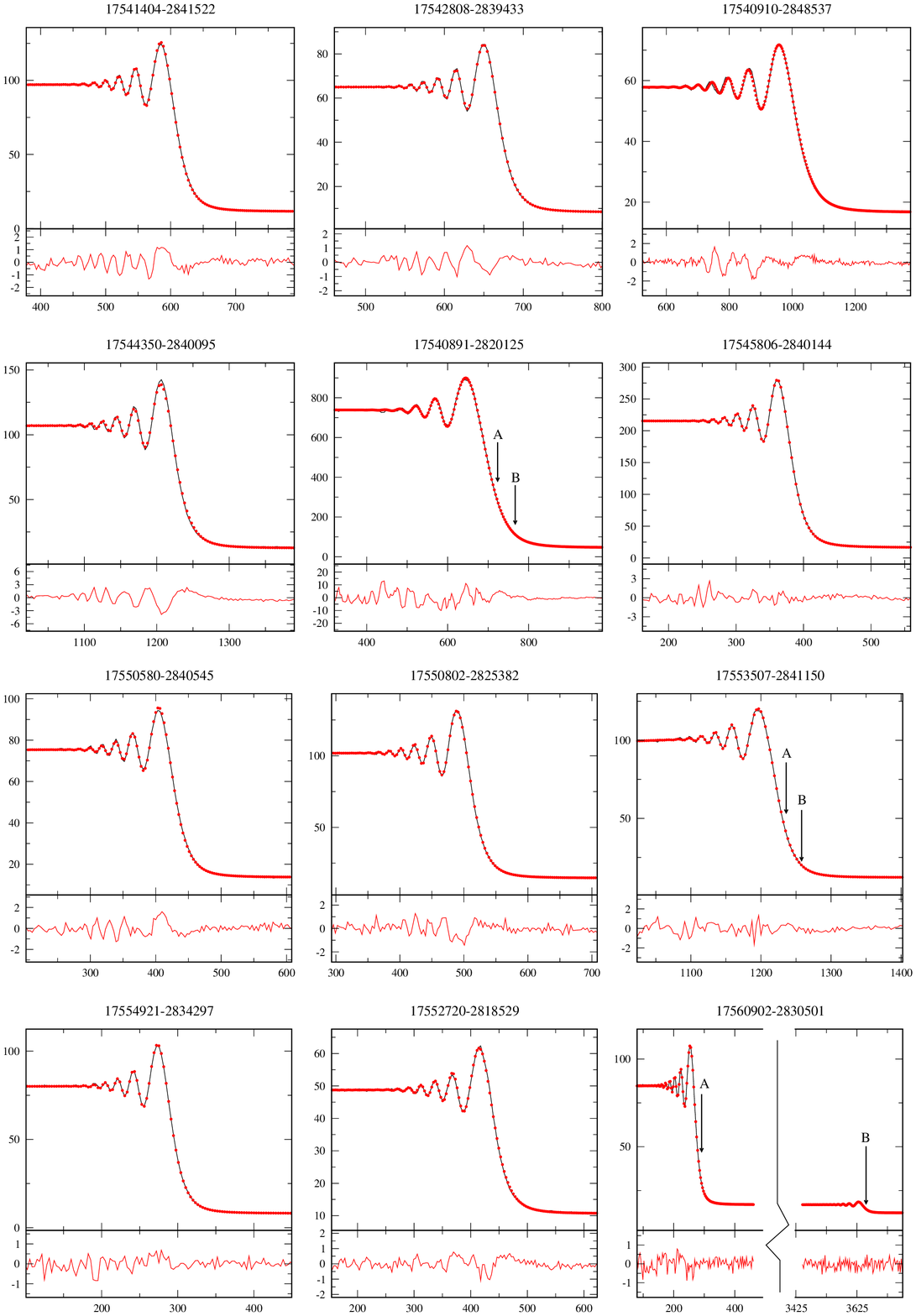}}
\caption {Top panels: LO data and
best fit. Bottom panels: fit residuals on an enlarged scale
(continued).}
\end{figure*}
}

%------------------------------------------------------------------------------
\onlfig{8}{
\begin{figure*}%f4
\resizebox{\hsize}{!}{\includegraphics{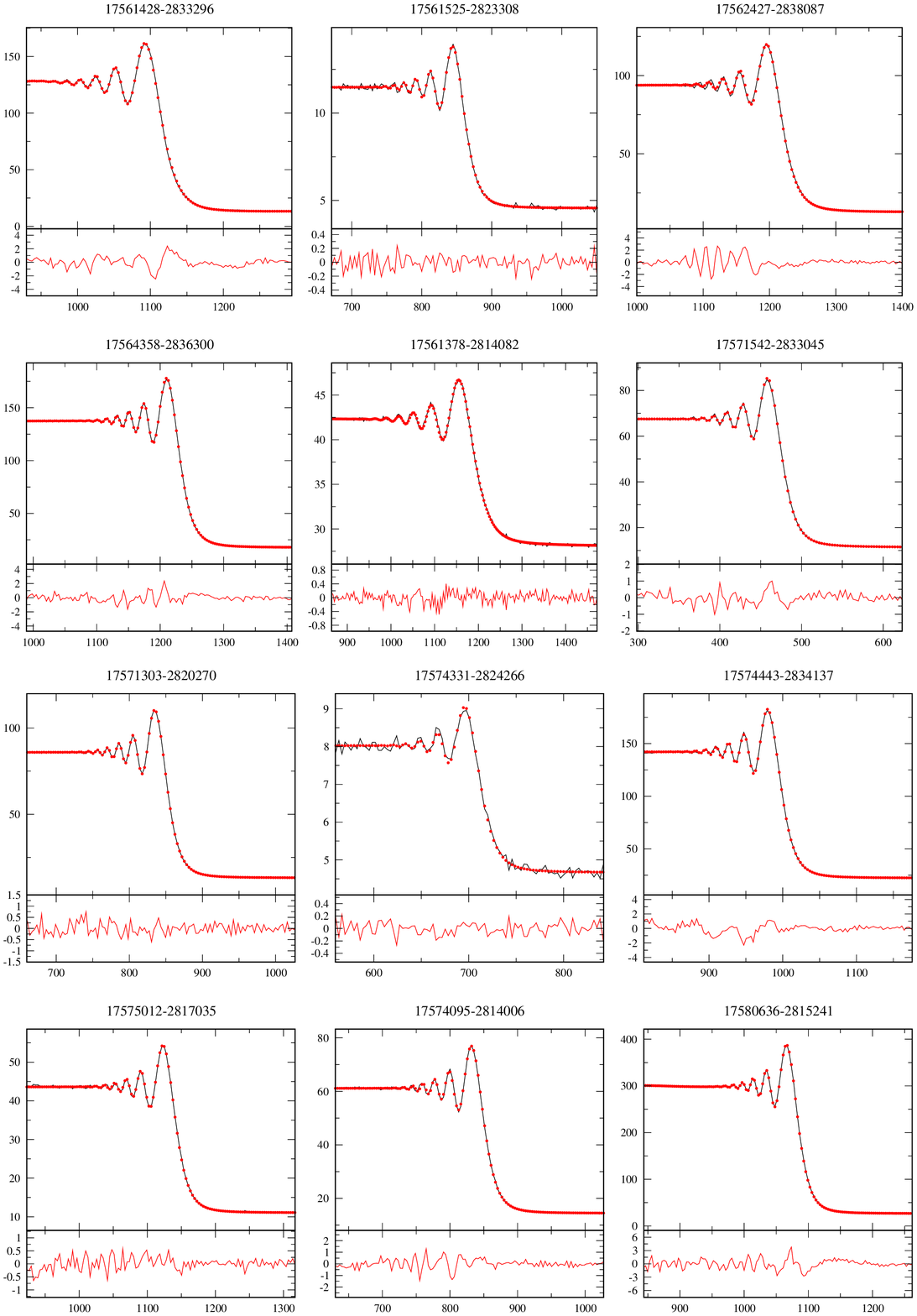}}
\caption {Top panels: LO data and
best fit. Bottom panels: fit residuals on an enlarged scale
(continued).}
\end{figure*}
}

%------------------------------------------------------------------------------
\onlfig{8}{
\begin{figure*}%f4
\resizebox{\hsize}{!}{\includegraphics{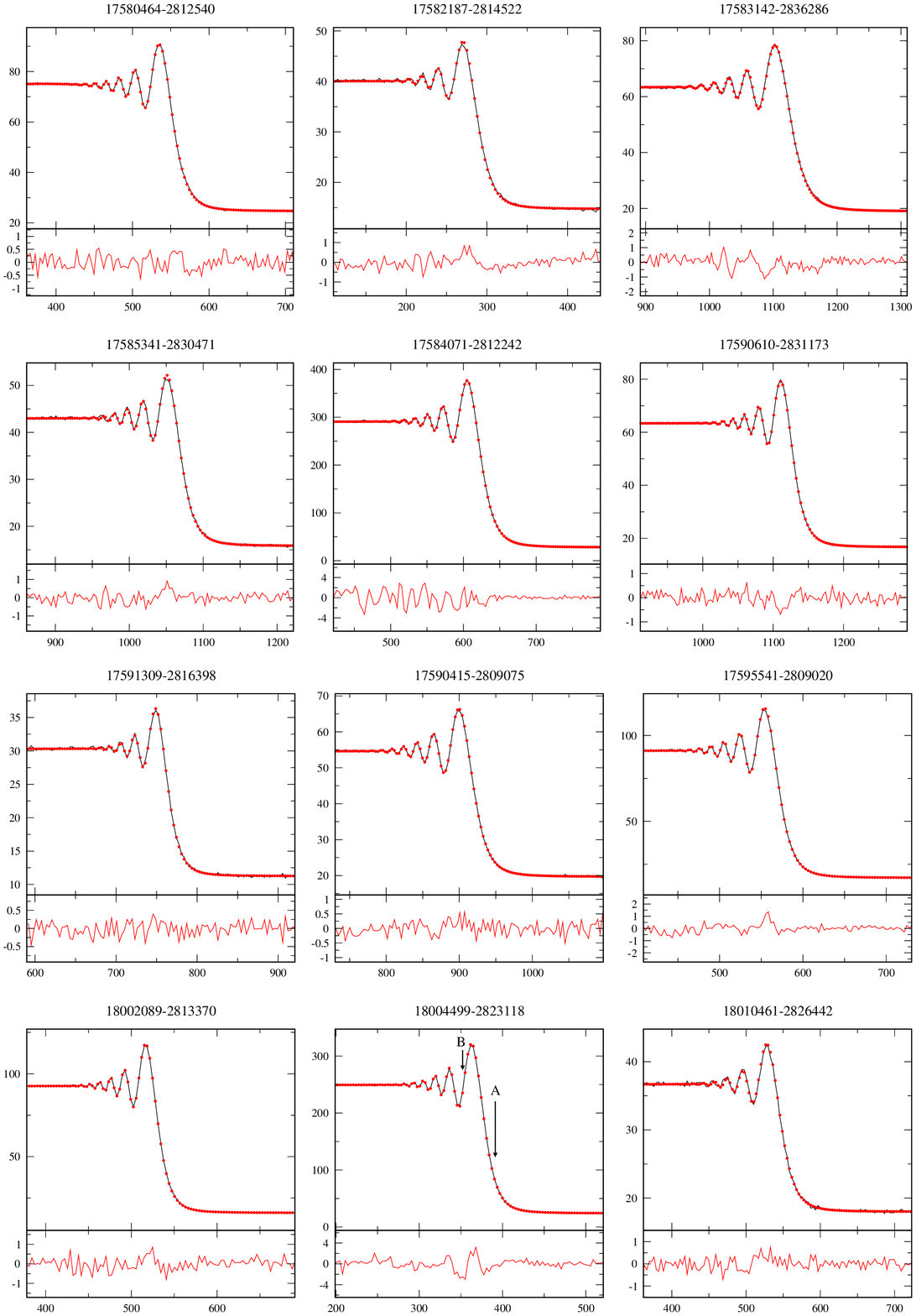}}
\caption {Top panels: LO data and
best fit. Bottom panels: fit residuals on an enlarged scale
(continued).}
\end{figure*}
}

%------------------------------------------------------------------------------
%------------------------------------------------------------------------------

%______________________________________________________________

\section{Conclusions}
We have reported on additional lunar occultation
observations, after those reported
in Paper~I, using the burst mode of the ISAAC instrument
at the ESO VLT. During a passage of the Moon in the central
regions of the galactic bulge, we have recorded 78 LO
events (72 with a positive detection)
over 8.5 hours. Our results include the new detection of
six binaries and one triple star, with 
typical projected separations of $\approx 10$\,mas,
i.e. five times less than the diffraction limit of the 8.2\,m
telescope. Brightness ratios were as low as 1:100.
We have also detected and measured the extended emission arising
from circumstellar dust around one maser star, two carbon stars,
the central star of a planetary nebula and other IR stars.

Our sources have no or very few known counterparts,
and other information such as fluxes or spectra is
also missing or very scarce.
Follow-up studies are necessary in order to better characterize
the energy distribution and spectral characteristics of these
sources. Our measurements have also shown a significant amount
of variability in many of the sources, with differences
between our determination and the 2MASS flux which approached
2\,mag in some cases.

As shown in this paper and in Paper~I,
the LO technique using a fast readout on a detector
subwindow at a large telescope  provides an unprecedented
combination of sensitivity and high angular resolution,
namely K$\approx$12\,mag and $0\farcs001$ respectively.
Fast-photometry on a detector subwindow
is now a well-characterized observational mode,
publicly offered at the VLT. 
%______________________________________________________________

\begin{acknowledgements}
This work is partially supported by the 
\emph{ESO Director General's Discretionary Fund}
and by the
\emph{MCYT-SEPCYT Plan Nacional I+D+I AYA \#2005-08604}.
AR wishes to thank the National Astronomical Observatory of
Japan in Mitaka, where he worked on the 
writing of this paper.
This research has made use of the SIMBAD database,
operated at CDS, Strasbourg, France.
\end{acknowledgements}

%_____________________________________________________________
%                              Table longer than a single page  
%  In the preamble, use:              \usepackage{longtable}
%-------------------------------------------------------------
%          All long tables have to be placed at the end, after 
%                                        \end{thebibliography}
%

% if table 2
\longtab{1}{
\begin{longtable}{lccrrrlll}
\caption{\label{lo_complete}
List of the recorded occultation events}\\
\hline\hline
\multicolumn{1}{c}{2MASS id}&
Frame&
NDIT&
\multicolumn{1}{c}{Time}&
\multicolumn{1}{c}{K}&
\multicolumn{1}{c}{J-K}&
\multicolumn{1}{c}{Sp}&
\multicolumn{1}{c}{Cross-Id} &Notes\\
\multicolumn{1}{c}{}&
(px)&
&
\multicolumn{1}{c}{(UT)}&
\multicolumn{2}{c}{(mag)}&
%\multicolumn{1}{c}{}&
\multicolumn{1}{c}{}&
\multicolumn{1}{c}{}&
\multicolumn{1}{c}{}\\
\hline
\endfirsthead
\caption{continued.}\\
\hline\hline
\multicolumn{1}{c}{2MASS id}&
Frame&
NDIT&
\multicolumn{1}{c}{Time}&
\multicolumn{1}{c}{K}&
\multicolumn{1}{c}{J-K}&
\multicolumn{1}{c}{Sp}&
\multicolumn{1}{c}{Cross-Id} &Notes\\
\multicolumn{1}{c}{}&
(px)&
&
\multicolumn{1}{c}{(UT)}&
\multicolumn{2}{c}{(mag)}&
%\multicolumn{1}{c}{}&
\multicolumn{1}{c}{}&
\multicolumn{1}{c}{}&
\multicolumn{1}{c}{}\\
\hline
\endhead
\hline
\endfoot
17455299-2833479	&	64	&	14000	&	22:32:02	&	8.43	&	5.89	&	\multicolumn{1}{c}{$-$}	&	\multicolumn{1}{c}{$-$}	&	not seen	\\
17461267-2848121	&	32	&	14000	&	23:11:38	&	6.45	&	4.93	&	\multicolumn{1}{c}{$-$}	&	\multicolumn{1}{c}{$-$}	&	not seen	\\
17472855-2825563	&	32	&	14000	&	23:29:29	&	7.25	&	8.29	&	\multicolumn{1}{c}{$-$}	&	\multicolumn{1}{c}{$-$}	&	piq	\\
17474895-2835083	&	32	&	14000	&	23:35:45	&	4.44	&	1.91	&	\multicolumn{1}{c}{$-$}	&	\multicolumn{1}{c}{$-$}	&	piq	\\
17475266-2842527	&	32	&	14000	&	23:43:36	&	5.62	&	1.56	&	\multicolumn{1}{c}{$-$}	&	\multicolumn{1}{c}{$-$}	&	piq	\\
17474310-2850553	&	64	&	12000	&	0:00:20	&	5.52	&	3.12	&	\multicolumn{1}{c}{$-$}	&	\multicolumn{1}{c}{$-$}	&		\\
17491814-2839483	&	32	&	14000	&	0:30:19	&	7.67	&	5.67	&	\multicolumn{1}{c}{$-$}	&	\multicolumn{1}{c}{$-$}	&	piq	\\
17492231-2830324	&	32	&	14000	&	0:37:02	&	4.64	&	1.87	&	\multicolumn{1}{c}{$-$}	&	\multicolumn{1}{c}{$-$}	&		\\
17492803-2845256	&	32	&	14000	&	0:41:27	&	4.77	&	1.84	&	M6	&	2MASS J17492803-2845256	&		\\
17494120-2836360	&	32	&	14000	&	0:44:20	&	5.53	&	1.64	&	\multicolumn{1}{c}{$-$}	&	\multicolumn{1}{c}{$-$}	&	soc\\
17495464-2839594	&	32	&	14000	&	0:52:57	&	5.96	&	4.17	&	\multicolumn{1}{c}{$-$}	&	\multicolumn{1}{c}{$-$}	&		\\
17491339-2823089	&	32	&	12000	&	0:56:40	&	7.55	&	6.34	&	\multicolumn{1}{c}{$-$}	&	\multicolumn{1}{c}{$-$}	&		\\
17500552-2839384	&	32	&	14000	&	0:59:40	&	5.83	&	3.08	&	\multicolumn{1}{c}{$-$}	&	\multicolumn{1}{c}{$-$}	&		\\
17500584-2844582	&	32	&	6904	&	1:03:42	&	6.05	&	3.04	&	\multicolumn{1}{c}{$-$}	&	\multicolumn{1}{c}{$-$}	&	not seen	\\
17500960-2829492	&	32	&	14000	&	1:09:32	&	5.47	&	1.49	&	\multicolumn{1}{c}{$-$}	&	\multicolumn{1}{c}{$-$}	&	soc\\
17502419-2843494	&	64	&	8000	&	1:13:48	&	8.36	&	5.78	&	\multicolumn{1}{c}{$-$}	&	\multicolumn{1}{c}{$-$}	&		\\
17503131-2839110	&	32	&	2190	&	1:15:53	&	6.82	&	3.83	&	\multicolumn{1}{c}{$-$}	&	\multicolumn{1}{c}{$-$}	&	not seen	\\
17501676-2827010	&	32	&	14000	&	1:21:22	&	4.97	&	1.39	&	\multicolumn{1}{c}{$-$}	&	\multicolumn{1}{c}{$-$}	&		\\
17503361-2827556	&	32	&	14000	&	1:29:47	&	6.94	&	3.30	&	\multicolumn{1}{c}{$-$}	&	\multicolumn{1}{c}{$-$}	&		\\
17510350-2839348	&	32	&	14000	&	1:36:23	&	5.31	&	0.87	&	\multicolumn{1}{c}{$-$}	&	\multicolumn{1}{c}{$-$}	&		\\
17505456-2828313	&	32	&	12000	&	1:41:49	&	6.22	&	2.50	&	\multicolumn{1}{c}{$-$}	&	\multicolumn{1}{c}{$-$}	&		\\
17505555-2850170	&	32	&	14000	&	1:44:38	&	5.71	&	2.33	&	\multicolumn{1}{c}{$-$}	&	\multicolumn{1}{c}{$-$}	&	edge	\\
17510869-2829380	&	32	&	14000	&	1:48:21	&	5.03	&	2.71	&	\multicolumn{1}{c}{$-$}	&	\multicolumn{1}{c}{$-$}	&		\\
17513148-2841473	&	32	&	14000	&	1:54:50	&	5.19	&	2.31	&	\multicolumn{1}{c}{$-$}	&	\multicolumn{1}{c}{$-$}	&		\\
17514002-2840031	&	32	&	10000	&	1:59:39	&	3.91	&	1.46	&	\multicolumn{1}{c}{$-$}	&	\multicolumn{1}{c}{$-$}	&	not seen	\\
17505630-2823476	&	32	&	14000	&	2:02:33	&	5.34	&	1.21	&	\multicolumn{1}{c}{$-$}	&	\multicolumn{1}{c}{$-$}	&	soc\\
17515263-2837145	&	32	&	14000	&	2:07:47	&	5.76	&	1.51	&	\multicolumn{1}{c}{$-$}	&	\multicolumn{1}{c}{$-$}	&	soc\\
17512677-2825371	&	32	&	14000	&	2:11:15	&	5.97	&	3.28	&	\multicolumn{1}{c}{$-$}	&	GAL 001.095-00.832	&	long	\\
17514339-2825469	&	32	&	14000	&	2:20:34	&	4.54	&	1.40	&	\multicolumn{1}{c}{$-$}	&	\multicolumn{1}{c}{$-$}	&		\\
17522359-2840280	&	32	&	14000	&	2:27:17	&	4.66	&	1.48	&	\multicolumn{1}{c}{$-$}	&	\multicolumn{1}{c}{$-$}	&		\\
17521081-2827039	&	32	&	14000	&	2:32:43	&	3.96	&	1.08	&	M2	&	HD 316515	&		\\
17521791-2827103	&	32	&	8000	&	2:36:36	&	3.80	&	1.71	&	\multicolumn{1}{c}{$-$}	&	\multicolumn{1}{c}{$-$}	&		\\
17523994-2829496	&	32	&	14000	&	2:43:58	&	7.68	&	4.56	&	\multicolumn{1}{c}{$-$}	&	\multicolumn{1}{c}{$-$}	&		\\
17524687-2847207	&	32	&	14000	&	2:49:36	&	5.35	&	2.58	&	\multicolumn{1}{c}{$-$}	&	IRAS 17496-2846	&		\\
17530735-2837221	&	32	&	14000	&	2:54:08	&	4.99	&	1.39	&	\multicolumn{1}{c}{$-$}	&	\multicolumn{1}{c}{$-$}	&		\\
17532347-2842169	&	32	&	8000	&	3:05:50	&	5.95	&	3.08	&	\multicolumn{1}{c}{$-$}	&	\multicolumn{1}{c}{$-$}	&		\\
17524903-2822586	&	32	&	14000	&	3:08:10	&	5.21	&	1.33	&	\multicolumn{1}{c}{$-$}	&	\multicolumn{1}{c}{$-$}	&		\\
17531817-2849492	&	32	&	8000	&	3:17:43	&	4.30	&	2.64	&	\multicolumn{1}{c}{$-$}	&	C* 2490	&		\\
17534818-2841185	&	32	&	14000	&	3:20:22	&	5.35	&	0.65	&	K0III	&	HD 162761	&		\\
17540460-2837587	&	32	&	14000	&	3:28:41	&	4.93	&	1.30	&	\multicolumn{1}{c}{$-$}	&	\multicolumn{1}{c}{$-$}	&	2 stars	\\
17541283-2833521	&	32	&	10000	&	3:33:53	&	4.64	&	2.17	&	\multicolumn{1}{c}{$-$}	&	IRAS 17510-2833	&		\\
17541404-2841522	&	32	&	14000	&	3:36:48	&	5.42	&	2.31	&	\multicolumn{1}{c}{$-$}	&	\multicolumn{1}{c}{$-$}	&		\\
17542808-2839433	&	32	&	14000	&	3:43:29	&	5.53	&	1.56	&	\multicolumn{1}{c}{$-$}	&	\multicolumn{1}{c}{$-$}	&	piq	\\
17540910-2848537	&	32	&	10000	&	3:50:10	&	5.86	&	1.32	&	\multicolumn{1}{c}{$-$}	&	\multicolumn{1}{c}{$-$}	&		\\
17544350-2840095	&	32	&	14000	&	3:53:04	&	4.97	&	1.60	&	\multicolumn{1}{c}{$-$}	&	\multicolumn{1}{c}{$-$}	&		\\
17540891-2820125	&	32	&	10000	&	3:57:03	&	3.76	&	2.77	&	M8+:	&	NSV 9818	&		\\
17545806-2840144	&	32	&	14000	&	4:01:52	&	4.39	&	2.08	&	\multicolumn{1}{c}{$-$}	&	\multicolumn{1}{c}{$-$}	&		\\
17550580-2840545	&	32	&	14000	&	4:07:14	&	7.19	&	4.06	&	\multicolumn{1}{c}{$-$}	&	\multicolumn{1}{c}{$-$}	&	piq	\\
17550802-2825382	&	32	&	10000	&	4:11:16	&	5.00	&	2.19	&	\multicolumn{1}{c}{$-$}	&	\multicolumn{1}{c}{$-$}	&		\\
17553507-2841150	&	32	&	10000	&	4:25:23	&	4.11	&	2.27	&	\multicolumn{1}{c}{$-$}	&	IRAS 17524-2840	&		\\
17554921-2834297	&	32	&	14000	&	4:27:54	&	5.23	&	1.25	&	\multicolumn{1}{c}{$-$}	&	\multicolumn{1}{c}{$-$}	&		\\
17552720-2818529	&	32	&	12000	&	4:34:23	&	5.93	&	1.40	&	\multicolumn{1}{c}{$-$}	&	\multicolumn{1}{c}{$-$}	&		\\
17560902-2830501	&	32	&	12000	&	4:38:31	&	5.32	&	1.85	&	\multicolumn{1}{c}{$-$}	&	\multicolumn{1}{c}{$-$}	&	2stars	\\
17561428-2833296	&	32	&	14000	&	4:41:19	&	5.63	&	2.82	&	\multicolumn{1}{c}{$-$}	&	\multicolumn{1}{c}{$-$}	&	2stars	\\
17561525-2823308	&	32	&	12000	&	4:46:51	&	7.08	&	3.76	&	\multicolumn{1}{c}{$-$}	&	\multicolumn{1}{c}{$-$}	&		\\
17562427-2838087	&	32	&	14000	&	4:50:03	&	4.63	&	1.89	&	\multicolumn{1}{c}{$-$}	&	\multicolumn{1}{c}{$-$}	&		\\
17564358-2836300	&	32	&	14000	&	4:59:08	&	4.68	&	1.42	&	\multicolumn{1}{c}{$-$}	&	\multicolumn{1}{c}{$-$}	&		\\
17561378-2814082	&	32	&	12000	&	5:09:03	&	7.40	&	5.33	&	\multicolumn{1}{c}{$-$}	&	\multicolumn{1}{c}{$-$}	&		\\
17571542-2833045	&	32	&	14000	&	5:13:41	&	5.49	&	1.72	&	\multicolumn{1}{c}{$-$}	&	\multicolumn{1}{c}{$-$}	&		\\
17571303-2820270	&	32	&	14000	&	5:17:02	&	5.26	&	1.52	&	\multicolumn{1}{c}{$-$}	&	\multicolumn{1}{c}{$-$}	&		\\
17574331-2824266	&	32	&	10000	&	5:27:23	&	6.54	&	2.93	&	\multicolumn{1}{c}{$-$}	&	\multicolumn{1}{c}{$-$}	&		\\
17574443-2834137	&	32	&	14000	&	5:29:59	&	4.67	&	1.03	&	\multicolumn{1}{c}{$-$}	&	\multicolumn{1}{c}{$-$}	&		\\
17575952-2825406	&	32	&	10000	&	5:34:46	&	4.50	&	2.05	&	\multicolumn{1}{c}{$-$}	&	\multicolumn{1}{c}{$-$}	&	not seen	\\
17575012-2817035	&	32	&	8000	&	5:37:30	&	6.00	&	1.63	&	\multicolumn{1}{c}{$-$}	&	\multicolumn{1}{c}{$-$}	&		\\
17574095-2814006	&	32	&	14000	&	5:39:31	&	5.92	&	2.21	&	\multicolumn{1}{c}{$-$}	&	\multicolumn{1}{c}{$-$}	&		\\
17580636-2815241	&	32	&	12000	&	5:46:38	&	4.00	&	1.56	&	\multicolumn{1}{c}{$-$}	&	\multicolumn{1}{c}{$-$}	&		\\
17580464-2812540	&	32	&	10000	&	5:50:45	&	5.58	&	1.76	&	\multicolumn{1}{c}{$-$}	&	\multicolumn{1}{c}{$-$}	&		\\
17582187-2814522	&	32	&	14000	&	5:53:36	&	6.70	&	3.82	&	\multicolumn{1}{c}{$-$}	&	ESO 456-27	&		\\
17583142-2836286	&	32	&	8000	&	6:00:03	&	5.79	&	2.10	&	\multicolumn{1}{c}{$-$}	&	\multicolumn{1}{c}{$-$}	&		\\
17585341-2830471	&	32	&	12000	&	6:02:18	&	5.05	&	2.25	&	\multicolumn{1}{c}{$-$}	&	\multicolumn{1}{c}{$-$}	&		\\
17584071-2812242	&	32	&	10000	&	6:05:00	&	4.05	&	1.62	&	\multicolumn{1}{c}{$-$}	&	\multicolumn{1}{c}{$-$}	&		\\
17590610-2831173	&	32	&	8000	&	6:09:16	&	5.73	&	1.65	&	\multicolumn{1}{c}{$-$}	&	\multicolumn{1}{c}{$-$}	&		\\
17591309-2816398	&	32	&	14000	&	6:11:59	&	6.90	&	3.66	&	\multicolumn{1}{c}{$-$}	&	\multicolumn{1}{c}{$-$}	&		\\
17590415-2809075	&	32	&	14000	&	6:20:08	&	6.07	&	2.23	&	\multicolumn{1}{c}{$-$}	&	\multicolumn{1}{c}{$-$}	&		\\
17595541-2809020	&	32	&	12000	&	6:37:48	&	5.20	&	1.41	&	\multicolumn{1}{c}{$-$}	&	\multicolumn{1}{c}{$-$}	&		\\
18002089-2813370	&	32	&	14000	&	6:41:34	&	4.87	&	1.71	&	\multicolumn{1}{c}{$-$}	&	\multicolumn{1}{c}{$-$}	&		\\
18004499-2823118	&	32	&	14000	&	6:49:39	&	3.92	&	1.59	&	\multicolumn{1}{c}{$-$}	&	\multicolumn{1}{c}{$-$}	&		\\
18010461-2826442	&	32	&	10000	&	7:01:42	&	6.82	&	3.98	&	\multicolumn{1}{c}{$-$}	&	\multicolumn{1}{c}{$-$}	&	piq	\\
\end{longtable}
{\scriptsize
Frame is the size of the subwindow and NDIT the number of
frames used in the raw burst mode. The number
of frames in the reformatted FITS cube is NDIT/2-1, and this also
determines the total time span of the recorded data.
In the notes, piq stands for poor image quality and 
soc for slightly off center. Sometimes
2 stars were present in the field of view, but only one
was occulted.
}
}% End \longtab
\end{document}